\documentclass[aps,prl,twocolumn,superscriptaddress]{revtex4-1}
\usepackage{graphicx}

\begin{document}

\title{Magnetic Reconnection and Intermittent Turbulence in the Solar Wind}

\author{K.T. Osman}
\email{k.t.osman@warwick.ac.uk}
\affiliation{Centre for Fusion, Space and Astrophysics; University of Warwick, Coventry, CV4 7AL, United Kingdom}

\author{W.H. Matthaeus}
\affiliation{Bartol Research Institute, Department of Physics and Astronomy, University of Delaware, Delaware 19716, USA}

\author{J.T. Gosling}
\affiliation{Laboratory for Atmospheric and Space Physics, University of Colorado, Boulder, CO 80303, USA}

\author{A. Greco}
\author{S. Servidio}
\affiliation{Dipartimento di Fisica, Universit\`a della Calabria, 87036 Rende (CS), Italy}

\author{B. Hnat}
\affiliation{Centre for Fusion, Space and Astrophysics; University of Warwick, Coventry, CV4 7AL, United Kingdom}

\author{S.C. Chapman}
\affiliation{Centre for Fusion, Space and Astrophysics; University of Warwick, Coventry, CV4 7AL, United Kingdom}
\affiliation{Department of Mathematics and Statistics, University of Troms\o, N-9037 Troms\o, Norway}
\affiliation{Max Planck Institute for the Physics of Complex Systems, 01187 Dresden, Germany}

\author{T.D. Phan}
\affiliation{Space Sciences Laboratory, University of California, Berkeley, California 94720, USA}

\date{\today}

\begin{abstract}

A statistical relationship between magnetic reconnection, current sheets and intermittent turbulence in the solar wind is reported for the first time using \textit{in-situ} measurements from the Wind spacecraft at 1 AU. We identify intermittency as non-Gaussian fluctuations in increments of the magnetic field vector, $\mathbf{B}$, that are spatially and temporally non-uniform. The reconnection events and current sheets are found to be concentrated in intervals of intermittent turbulence, identified using the partial variance of increments method: within the most non-Gaussian 1\% of fluctuations in $\mathbf{B}$, we find 87\%--92\% of reconnection exhausts and $\sim$\,9\% of current sheets. Also, the likelihood that an identified current sheet will also correspond to a reconnection exhaust increases dramatically as the least intermittent fluctuations are removed from the dataset. Hence, the turbulent solar wind contains a hierarchy of intermittent magnetic field structures that are increasingly linked to current sheets, which in turn are progressively more likely to correspond to sites of magnetic reconnection. These results could have far reaching implications for laboratory and astrophysical plasmas where turbulence and magnetic reconnection are ubiquitous.

\end{abstract}

\pacs{}

\maketitle

\textit{Introduction}.---Turbulence is ubiquitous in space plasmas and leads to the emergence of coherent structures (see \citep{BrunoCarbone13} for a review). These are organized and concentrated structures such as current and vorticity sheets that are phase-correlated over their spatial extent, and are characterized by relatively long lifetimes. The solar wind is an ideal laboratory for the \textit{in-situ} study of coherent structures, which have traditionally been described as ideal magnetohydrodynamic (MHD) discontinuities \citep{Burlaga93}. However, these structures also display signatures of intermittency in the form of rare large amplitude magnetic and velocity field fluctuations that produce highly non-Gaussian heavy tailed probability distribution functions (PDF) \citep{Sorriso-ValvoEA99}, and have properties consistent with dynamical generation by strong plasma turbulence \citep{GrecoEA08,GrecoEA09,OwensEA11}. Indeed, turbulence generates coherent structures in hydrodynamics \citep{AnselmetEA84}, MHD \citep{MatthaeusMontgomery80,CarboneEA90,BiskampMuller00} and at kinetic scales in collisionless plasmas \citep{KarimabadiEA13}. Therefore, coherent structures embedded in the solar wind should reflect the nonlinear dynamics that give rise to intermittency \citep{Veltri99} such as random magnetic reconnection between adjoining flux tubes \citep{MatthaeusLamkin86, ServidioEA10,LoureiroEA09}. Here we ask whether some coherent structures in the solar wind might participate in magnetic reconnection, a fundamental process that converts magnetic energy into heat and plasma kinetic energy. This question, originally raised over 40 years ago \citep{BurlagaNess68recon}, is the subject of this Letter.

There has been renewed interest in the role of coherent magnetic structures in the solar wind based on the development and application of several identification techniques \citep{BrunoEA01,HadaEA03,VasquezEA07, GrecoEA08, LiEA11}. These intermittent structures are associated with enhanced turbulent dissipation in kinetic collisionless plasma simulations \citep{WanEA12,TenBargeHowes13} and non-uniform heating in the solar wind \citep{OsmanEA11,OsmanEA12a,WangEA13}. These results are consistent with the Kolmogorov refined similarity hypothesis \citep{Kolmogorov62} in neutral fluid turbulence theory, which relates local concentrations of the dissipation rate to large intermittent fluctuations. There is evidence to suggest solar wind proton temperature anisotropies are linked to coherent structures, which have been preferentially found in regions unstable with respect to plasma microinstabilities \citep{OsmanEA12b,ServidioEA14}. Recent work has also indicated these structures may contribute to the acceleration and transport of suprathermal particles \citep{TesseinEA13}. There is some observational support for the presence of coherent structures at sub-proton kinetic scales \citep{PerriEA12,WuEA13}. These can cause deviations from local thermal equilibrium in velocity distribution functions \citep{GrecoEA12, ServidioEA12, PerroneEA13,HaynesEA13}. Therefore, relationships must exist between coherent structures, intermittent turbulence, plasma heating and broader kinetic activity.

Magnetic reconnection is also an important element in a broad range of space and laboratory plasmas \citep{YamadaEA10}. These plasmas are often turbulent \citep{SundkvistEA07}, but the influence of turbulence on the reconnection process is not fully understood. The link between intermittent plasma turbulence and magnetic reconnection is well supported by numerical simulations \citep{MatthaeusVelli11}. However, similar statistical connections have not yet been found in the solar wind. This is due, at least in part, to a lack of sufficient numbers of identified reconnection events. While magnetic reconnection in the solar wind was originally thought to be relatively rare \citep{GoslingEA05, GoslingEA06a,GoslingEA06b,Gosling07}, higher resolution measurements coupled with refined techniques have recently led to an increase in the number of identified cases \citep{GoslingEA07,GoslingSzabo08,Gosling12}. This has now reached a point where a statistically meaningful study of the type presented here can be conducted. This Letter presents novel observational results linking magnetic reconnection to non-Gaussian features in the solar wind that are associated with intermittent turbulence. 

\textit{Analysis}.---We use 3 s resolution magnetic field measurements from the Magnetic Field Investigation (MFI) \citep{LeppingEA95} and proton moments from the 3D Plasma Analyzer (3DP) \citep{LinEA95} onboard the Wind spacecraft. The data intervals used in this investigation are listed in Table 1, and were originally selected randomly and then carefully scrutinized for reconnection exhausts. In the solar wind Petschek-like exhausts resulting from reconnection are identified as roughly Alfv\'enic-jetting plasma (based on the anti-parallel field components) that are bounded on one side by correlated changes in the anti-parallel components of $\mathbf{V}$ and $\mathbf{B}$ and by anti-correlated changes in those components on the other side. The list of identified reconnection exhausts that we use is assembled by application of these methods. In addition, current sheets are identified in a separate list as a reversal in at least one geocentric solar ecliptic (GSE) component of the magnetic field vector. While this method will miss current sheets where no field component actually reverses sign, it will not significantly affect our results since these are likely to be associated only with small fluctuations and not the most intermittent structures of interest. The number of reconnection exhausts (RE) and current sheets (CS) identified in this way are listed in Table 1.

\begin{table}[h]
\caption{All the data analyzed in this study, including the number of reconnection exhausts (RE) and current sheets (CS). The mean correlation time $\langle \tau_{c} \rangle$ was computed in sub-intervals of 6h duration, and then averaged for each interval.}
\begin{tabular}{| c | c | c | c | c |}
\hline
Interval & Duration (yyyy mm dd) & RE & CS & $\langle \tau_{c} \rangle$ (min) \\
\hline
1 & 2001, 01 01 -- 02 03 UT & 138 & 28438 & 30 \\
2 & 2006, 03 01 -- 04 01 UT & 125 & 29062 & 26 \\
3 & 2007, 04 01 -- 05 01 UT & 105 & 28131 & 25 \\
4 & 2007, 06 01 -- 07 01 UT & 153 & 26639 & 27 \\
\hline
\end{tabular}
\end{table}

Here we investigate whether magnetic reconnection and current sheets are related to the intermittent character of the turbulent solar wind. Note that not all reconnecting current sheets are thought to be associated with turbulence, including reconnection in the heliospheric current sheet and at the leading or trailing edges of interplanetary coronal mass ejections. However, some fraction of reconnection exhausts in the ambient solar wind might be linked to intermittent plasma turbulence. To this end, the partial variance of increments (PVI) method is used to find coherent non-Gaussian magnetic field structures:
\begin{equation}
PVI  = \frac{\left| \Delta\mathbf{B} \right|}{\sqrt{\langle \left| \Delta\mathbf{B} \right|^{2} \rangle}}
\label{eq:PVI}
\end{equation}
where $\Delta\mathbf{B}(t,\tau) = \mathbf{B}(t + \tau) - \mathbf{B}(t)$ is the magnetic field vector increment using a time lag $\tau$ and $\langle \cdots \rangle$ denotes an appropriate time average. All the PVI results presented here use a 30 min interval of averaging, which corresponds roughly to the timescale $\tau_{c}$ reported in Table 1, where solar wind turbulent fluctuations become uncorrelated \citep{MatthaeusEA05}. In order to match the data resolution used when identifying reconnection exhausts and current sheets, a time lag $\tau = 3$ s is selected. This corresponds to a plasma frame spatial separation when using Taylor's hypothesis ($\mathbf{r} = -\mathbf{V}\tau$) \citep{Taylor38} that is within the inertial range, slightly larger than the typical transition to kinetic scales \citep{LeamonEA98}. The PVI time series is constructed in a manner that is connected with familiar diagnostics of intermittency and has had previous success identifying reconnection sites in direct numerical MHD \citep{ServidioEA11b} and Hall MHD \citep{DonatoEA13} turbulence simulations. However, the focus here will be on the identification of reconnection exhausts rather than the reconnection sites themselves. Events are selected by imposing thresholds on the magnetic PVI time series, which leads to a hierarchy of coherent structure intensities. This threshold level $\lambda$ is applied by finding the PVI value \textit{above} which $\lambda$\% of the observations are contained and then removing all the lower PVI data. Thus, the smallest values of $\lambda$ correspond to the largest values of PVI, and the most intermittent structures. The transition from Gaussian to non-Gaussian magnetic field fluctuations occurs around PVI $= 3$ \citep{OsmanEA11,OsmanEA12a}, and thus higher values of PVI ($\lambda \leq 1$) largely correspond to the non-Gaussian heavy tails of the magnetic field increment distribution. 

\textit{Results}.---Figure 1 shows selected plasma and magnetic field data for a 5 min interval encompassing a reconnection event. This exhaust was swept past the spacecraft in 9 s, and had a maximum local width of $3.3 \times 10^{3}$ km. It was also associated with a magnetic field rotation of $53^{\circ}$ that had a double-step character, which is characteristic of reconnection exhausts in the solar wind. The changes in $\mathbf{V}$ and $\mathbf{B}$ are anti-correlated on the leading edge of the exhaust and correlated on the trailing edge. These coupled changes correspond to Alfv\'enic disturbances that respectively propagated parallel and anti-parallel to $\mathbf{B}$ on reconnected field lines and bifurcated the original current sheet. The proton density and temperature were slightly enhanced within the exhaust and $V_{x}$ was slightly elevated, indicating the reconnection exhaust was directed sunward. This quasi-stationary magnetic reconnection event coincides with a significant increase in PVI, which shows a clear signature within the exhaust. Note that we cannot describe this interval as typical since there is considerable variation across all 521 reconnection events used in this study. However, the connection between elevated PVI and reconnection exhausts in the solar wind persists for most events in our statistical ensemble. 

\begin{figure}[h]
\includegraphics[width=8.5cm]{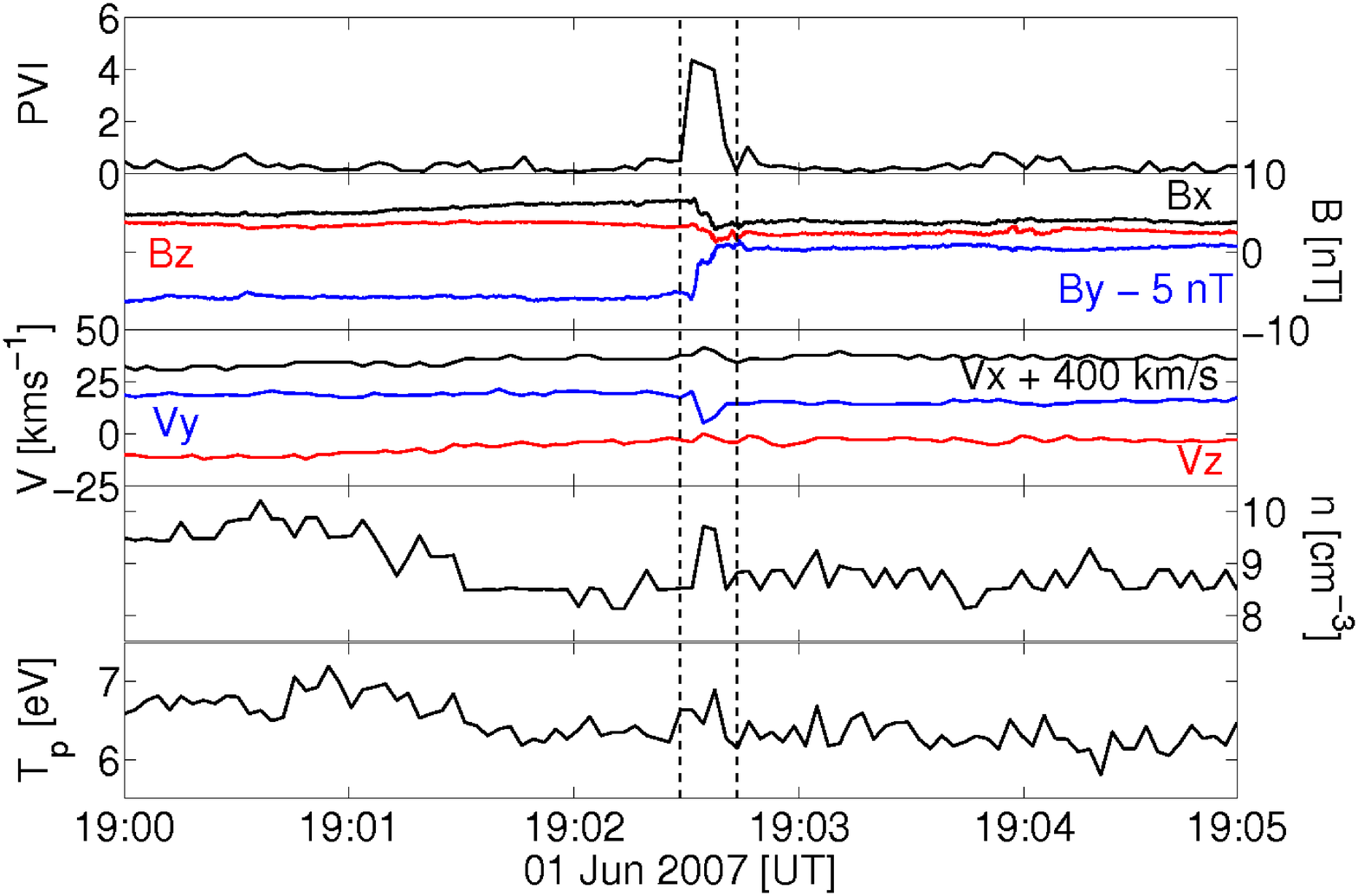}
\caption{An interval encompassing a reconnection exhaust. From top to bottom, the parameters plotted are PVI, the GSE components of the magnetic field and solar wind velocity, the proton number density and the proton temperature. In order to resolve the bifurcated current sheet, 92 ms resolution magnetic field data is used. All other plotted data has a resolution of 3 s. The x-component of the flow velocity has been shifted up by 400 kms$^{-1}$ and the y-component of the magnetic field has been shifted down by 5 nT. Vertical lines bracket the reconnection exhaust.}
\label{Fig:fig2}
\end{figure}

Table 2 lists the percentage of reconnection exhausts and current sheets identified in the solar wind for a selection of PVI thresholds. The application of these thresholds results in the exclusion of all but $\lambda$\% of the original dataset, leaving only the highest PVI values remaining. As the threshold is lowered, exhausts and current sheets associated with smaller PVI values are systematically removed. However, this response is not linear and some identified events remain even at the lowest thresholds. For example, 87--92\% of all reconnection exhausts and about 9\% of all current sheets are concentrated within the highest 1\% of PVI values. Therefore, the effective concentration of current sheets and Petschek-like exhausts is significantly increased by using PVI thresholds as data acceptance criteria. This trend exists at all PVI thresholds and for all four intervals used in this study. 

\begin{table}[h]
\caption{Percentage of magnetic reconnection events ($E_{RE}$) and current sheets ($E_{CS}$) identified at each PVI threshold ($\lambda$). For completeness, the average PVI value $\zeta$ that corresponds to the threshold is included (PVI $\geq \zeta$).}
\begin{tabular}{| c | c | c c | c c | c c | c c |}
\hline
&  & \multicolumn{2}{|c|}{Int. 1 (\%)} & \multicolumn{2}{|c|}{Int. 2 (\%)} & \multicolumn{2}{|c|}{Int. 3 (\%)} & \multicolumn{2}{|c|}{Int. 4 (\%)} \\ \cline{3-10}
\raisebox{2ex}{$\lambda$ (\%)} & \raisebox{2ex}{$\zeta$} & $E_{RE}$ & $E_{CS}$ & $E_{RE}$ & $E_{CS}$ & $E_{RE}$ & $E_{CS}$ & $E_{RE}$ & $E_{CS}$ \\
\hline
100          & 0      & 100    & 100    & 100  & 100   & 100  & 100    & 100  & 100 \\
   50         & 0.5   & 100    & 86.0   & 100  & 86.2    & 100  & 86.4   & 100  & 86.0 \\
   10         & 1.4   & 100    & 40.7   & 100  & 41.4   & 100  & 41.5   & 100  & 40.6 \\
    5          & 2.0   &   99.3 & 27.0   & 100  & 27.8   & 100  & 27.7   & 98.0 & 26.2 \\
    1          & 3.5   &   87.7 & 9.4   & 90.4 & 9.6   & 92.4 & 9.7   & 86.9 & 9.0 \\
    0.5       & 4.4   &   72.5 & 5.7   & 76.0 & 5.9   & 81.9 & 5.9   & 79.1 & 5.5 \\
    0.1       & 6.6   &   29.0 & 1.6   & 36.8 & 1.6  & 41.9 & 1.7   & 44.4 & 1.5 \\
    0.05     & 7.6   &   22.5 & 0.9     & 26.4 & 0.9     & 25.7 & 0.9      & 32.7 & 0.8  \\
    0.01     & 10.2 &    9.4  & 0.2    & 11.2 & 0.2     & 13.3 & 0.2      & 13.1 & 0.2      \\
    0.005   & 11.2 &    6.5  & 0.1     & 7.2   & 0.1     & 6.7    & 0.1      & 7.8   & 0.1     \\
    0.001   & 14.4 &    2.2  & 0.02   & 2.4   & 0.03   & 2.9    & 0.02   & 3.3   & 0.02   \\
    0.0005 & 15.8 &   1.4   & 0.007 & 0.8   & 0.01   & 1.9    & 0.01   & 2.0   & 0.01 \\
    0.0001 & 18.1 &   0.7      & 0.004         & 0.8   & 0.003 & 1.0    & 0.004 & 0.7   & 0.004  \\
\hline
\end{tabular}
\end{table}

Figure 2 shows the percentage of reconnection exhausts and current sheets at each PVI threshold normalized to the percentage of data occupied by these events. This concentration of reconnection events $C_{RE} = E_{RE}/\lambda$ and current sheets $C_{CS} = E_{CS}/\lambda$ should remain around unity for every threshold if there is no PVI dependence. However, it is clear that both reconnection events and current sheets are concentrated preferentially at the largest PVI values, which are related to the intermittent properties of turbulence. Note that reconnection exhausts are significantly more concentrated than current sheets at all but the largest $\lambda$ thresholds. While the dependence of current sheet concentration on PVI is to be expected, the more intermittent character of the exhausts is a novel result. 

The diverging behavior of magnetic reconnection and current sheet concentrations in Fig. 2 suggests that the relationship between these events may depend upon PVI threshold. Figure 3 shows the percentage of current sheets that correspond to reconnection exhausts at each threshold, $(RE/CS)\times(E_{RE}/E_{CS}) \times 100$. For the highest thresholds (lowest PVI), the number of non-reconnecting current sheets far exceeds the number of exhausts. As the threshold decreases, many more current sheets are removed in comparison to reconnection events. The percentage of reconnecting current sheets increases from around 0.5\% in the entire original dataset to eventually reaching 100\% for the highest PVI values. Therefore, this suggests the turbulent solar wind has a hierarchy of intermittent structures that are increasingly linked to current sheets, which in turn are more likely to correspond to sites of magnetic reconnection.

\begin{figure}[h]
\includegraphics[width=8.5cm]{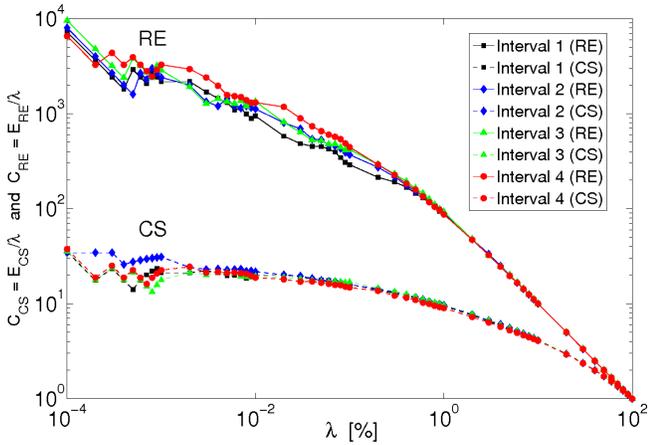}
\caption{The concentration of magnetic reconnection exhausts $C_{RE} = E_{RE}/\lambda$ and current sheets $C_{CS} = E_{CS}/\lambda$ at each PVI threshold $\lambda$. Each color corresponds to a different interval listed in Table 1.}
\label{Fig:fig3}
\end{figure}

\textit{Discussion}.---We have presented the first direct evidence that Petschek-like exhausts in the solar wind are statistically associated with non-Gaussian, large amplitude magnetic field fluctuations, which are thought to be connected to the intermittent character of MHD turbulence. This result provides further insight into the relationship between these fundamental processes. However, the exact nature of this link between magnetic reconnection and plasma turbulence is unclear. It is known that MHD turbulence dynamically generates current sheet-like coherent structures that are candidate sites for active reconnection \citep{MatthaeusMontgomery80, MatthaeusLamkin86,ServidioEA10}. Thus, the reconnection exhausts seen in the solar wind could result from reconnection sites that are dynamically generated by plasma turbulence. Alternatively, intermittent turbulence could be driven by reconnection exhausts \citep{Lapenta08}. While there is not yet any direct observational evidence for turbulence generated by reconnection exhausts in the solar wind, MHD simulations \citep{MatthaeusLamkin86} have suggested that there is a complicated feedback mechanism underlying the interaction between plasma turbulence and magnetic reconnection. Indeed, turbulence has been observed within a magnetic reconnection ion diffusion region in the magnetotail \citep{EastwoodEA09}, and kinetic simulations suggest that fluctuations generated by magnetic reconnection can exhibit the hallmarks of intermittent turbulence \citep{LeonardisEA13}. These results could have far reaching implications in laboratory and astrophysical plasmas where turbulence and magnetic reconnection are ubiquitous. 

\begin{figure}[h]
\includegraphics[width=8.5cm]{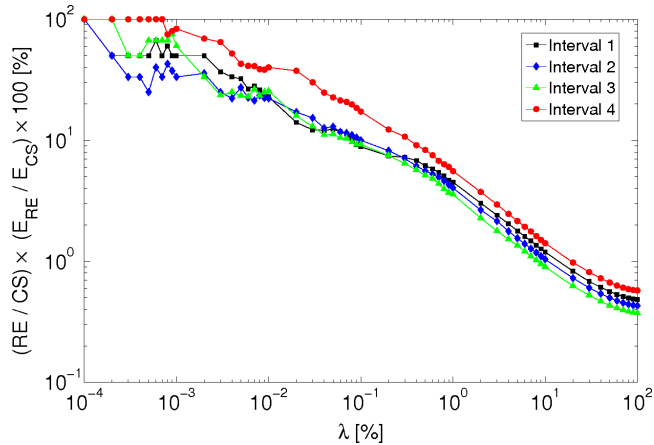}
\caption{The percentage of current sheets that correspond to magnetic reconnection exhausts $[(RE/CS)\times(E_{RE}/E_{CS})\times100]$ at each PVI threshold $\lambda$. Each color represents a different interval listed in Table 1.}
\label{Fig:fig4}
\end{figure}

The present study has demonstrated that application of increasing PVI thresholds on solar wind data acts to increase, within the selected population, the concentration of current sheets and reconnection exhausts. It also increases the probability that an identified current sheet will correspond to a reconnection event. Therefore, the PVI method could form the basis of an automated reconnection identification tool, and has previously been successful in numerical simulations \citep{ServidioEA11b}. At the highest thresholds the PVI statistic identified all reconnection exhausts, but these are greatly outnumbered by other non-reconnection events (false positives). As the threshold is lowered a greater percentage of the remaining events correspond to reconnection exhausts, but increasing numbers of exhausts are not identified. From this perspective the PVI method cannot supplant more detailed exhaust identification methodologies. However, its simplicity and exclusive reliance upon magnetic field measurements, which are typically available at a higher resolution than plasma measurements, make it attractive in some practical applications. This includes, but is not limited to, burst-mode triggers on spacecraft instruments.

Further work is required to determine whether the relationship between PVI and magnetic reconnection exhausts in the solar wind is universal. The overall physical nature and occurrence rate of reconnecting current sheets depends on solar wind speed; high speed streams contain current sheets that are more Alfv\'enic and fewer in number than those found in low speed streams \citep{GoslingEA06a,GoslingEA06b}. Note that the PVI method used here does not explicitly distinguish between fast and slow speed solar wind. It also ignores the effects of shear angle and proton plasma beta, combinations of which can suppress magnetic reconnection even at thin current sheets \citep{SwisdakEA10,PhanEA10,GoslingPhan13}. However, solar wind speed and plasma beta can also affect the nature of MHD turbulence \citep{PetrosyanEA10}, and thus the associated PVI values could be modified in different parameter regimes. Hence, similar studies will be conducted in different solar wind streams and plasma beta regimes with the aim of reproducing the present results. In addition, work has already begun on investigating the interaction between MHD turbulence, magnetic reconnection and kinetic effects such as temperature anisotropy.

Laboratory and space plasma observations suggest fast reconnection onset occurs when the current layer approaches ion Larmor scales \citep{VaivadsEA04,EgedalEA07}. In addition, plasma turbulence generates current sheets on all dynamical scales from the proton to electron gyroradius, and these have been observed in the solar wind \citep{PerriEA12} and direct numerical simulations \citep{KarimabadiEA13,HaynesEA13}. However, magnetic reconnection in the solar wind often does not lead directly to dissipation and plasma heating, particularly for asymmetric boundary conditions and modest shear angles \citep{Gosling12}. This could be because the highest resolution combined plasma and magnetic field \textit{in-situ} data currently available to identify reconnection exhausts in the solar wind is generally insufficient to access the relevant kinetic scales. The upcoming NASA Deep Space Climate Observatory (DSCOVR) and Magnetospheric Multi-Scale (MMS) missions will make high-resolution plasma and magnetic field measurements, and thus should identify more reconnection events across a wide range of spatial scales. This will further improve our understanding of magnetic reconnection and turbulence at energy dissipation scales, and could have far reaching implications for turbulent dissipation in collisionless plasmas. 

This research is supported by UK STFC, EU Turboplasmas project (Marie Curie FP7 PIRSES-2010-269297), NASA Magnetospheric Multi-Scale Mission Theory and Modeling program (NNX08AT76G) and Solar Probe Plus ISIS project, NASA grants NNX08AO84G, NNX10AF26G and NNX08AO83G, and NSF SHINE (AGS-1156094) and Solar Terrestrial (AGS-1063439) programs, and in Italy by POR Calabria FSE 2007/2013.

\end{document}